\documentclass{aa}
\usepackage{graphicx,epsf,epsfig,float}
\usepackage{amssymb} 

\def\ion#1#2{{\rm #1}~{\sc #2}}

\def\agata{R\' o\. za\' nska}

\def\NH{$N_{\rm H}$}
\def\nh{$n_{\rm H}$}

\def\titan{{\sc TITAN}}

\begin{document}

\title{Thermal instability in X-ray photoionized media \\ 
in Active Galactic Nuclei:} 
\subtitle{Influence on the gas structure and spectral features}

\author{A. C. Gon\c{c}alves
        \inst{1,2} 
        \and 
	S. Collin\inst{1}
	\and
	A.-M. Dumont\inst{1}
	\and
	L. Chevallier\inst{1,3}
         }

\offprints{A. C. Gon\c{c}alves}

\institute{
LUTH, Observatoire de Paris, Section de
Meudon, 5 Place Jules Janssen, 92195 Meudon Cedex, France \\
(\email{anabela.goncalves@obspm.fr}) 
\and
CAAUL, Observat\'orio Astron\'omico de Lisboa,
Tapada da Ajuda, 1349-018 Lisboa, Portugal
\and
Nicolaus Copernicus Astronomical Center,
Bartycka 18, 00-716 Warszawa, Poland
          }
   \date{Received 21 June 2006 / Accepted 28 Novembre 2006}

\titlerunning{Thermal instability in X-ray photoionized media in AGN}
\authorrunning{ A. C. Gon\c{c}alves et al.}

\abstract
    {A photoionized gas in thermal equilibrium can 
display a thermal instability, with three or more 
solutions in the multi-branch region of the S-shape curve giving 
the temperature versus the radiation-to-gas-pressure 
ratio. Many studies have 
been devoted to this curve and to its dependence on 
different parameters, always in the optically thin case.}
   {The purpose of our study is the 
thermal instability in 
optically thick, stratified media, in  
total pressure equilibrium. We are also 
interested in comparing photoionization models 
issued from the hot and cold stable solutions,  
with the currently used models computed 
with an approximate, intermediate solution. 
}  
   {We have developped a new algorithm to select 
the hot/cold stable solution,  
and thereof to compute a fully consistent photoionization 
model. We have implemented it in the TITAN code and computed 
a set of models encompassing the range of conditions 
valid for the Warm Absorber in Active 
Galactic Nuclei. 
}
   {We have demonstrated that the thermal 
instability problem is quite different in thin or thick media.   
Models computed with the hot/cold stable solutions, or with an 
intermediate solution, differ  all along the gas slab, 
with the spectral distribution changing as 
the radiation progresses inside the ionized gas.   
These effects depend on the thickess of the medium and 
on its ionization.} 
   {This has observational implications in the emitted/absorbed 
spectra, ionization states, and variability. However  
impossible to know what solution the plasma 
will adopt when attaining the multi-solutions regime,   
we expect the  emitted/absorbed spectrum  
to be intermediate between those resulting from pure cold 
and hot models; such a phase-mixed medium can be well 
reproduced by intermediate solution models.  
Large spectral fluctuations corresponding to the 
onset of a cold/hot solution could be observed in 
timescales of the order of the dynamical time. A 
strong turbulence implying supersonic velocities 
should permanently exist in the multi-branch region of thick, 
stratified, pressure equilibrium media.	
}

\maketitle

\section{Introduction}

It is well known that a photoionized gas in thermal equilibrium, 
i.e. a gas where the radiative heating is balanced by radiative 
cooling, can display a thermal instability (e.g., Krolik, Tarter \& McKee 
1981). The phenomenon manifests itself in the S-shape of the net 
cooling function\footnote{In this study, the net cooling 
function ${\cal L}$ is defined as the difference between 
the cooling ($\Lambda$) and heating ($\Gamma$) functions in a 
given region, and is expressed in units of erg\,cm$^{3}$\,s$^{-1}$. 
It is also possible to use a net cooling function 
describing the energy balance per gram of material, per second.} 
or, which is equivalent, through the 
curve giving the temperature versus the radiation-to-gas-pressure
ratio. Such an S-shape curve allows for the co-existence of gas 
at different temperatures and densities for the same pressure ratio.  

For a given value of the radiation-to-gas-pressure ratio, the gas 
can then be in three (or even more) 
states of thermal equilibrium, which depend on the Spectral Energy 
Distribution (SED) of the ionizing spectrum,  on the abundances, 
and on other physical  parameters affecting the heating and the 
cooling. One of these states is thermally unstable, as it does not 
satisfy the stability criterium to isobaric perturbations (Field
1965): 
\begin{equation}
(\frac{\partial {\cal L}}{\partial T})_{P_{\rm gas}} > 0,
\label{eq-field}
\end{equation}
where  ${\cal L}$ is the net cooling function, $T$ 
is the temperature, and $P_{\rm gas}$ 
is the gas pressure.  The two other states for which the 
derivative of ${\cal L}$ is positive, are stable. 
Such thermal instability is, for instance, at the origin 
of the two-phase model for the interstellar medium, where 
cold ($T\lesssim 300$~K) neutral atomic and molecular clouds 
are embedded in a warm ($T\sim 10^{4}$~K) intercloud medium 
(Field, Goldsmith \& Habing 1969). 

\subsection{Optically thin media}
Krolik et al. (1981) studied the thermal instability in the 
context of Active Galactic Nuclei (AGN), for {\it optically thin},  
X-ray illuminated gas. In such media, the radiation 
pressure keeps a constant value, and the 
energy balance equation solution depends on the gas 
pressure only. When the gas pressure is small compared 
to the radiation pressure, there is a unique, stable, ``hot''  
solution, where both the heating and the cooling are 
dominated by Compton processes (Compton heating, 
inverse Compton cooling). As the gas pressure 
increases, atomic processes (photoionization heating,  
line and continuum cooling) become important, and multiple 
solutions arise. The harder the spectrum, the more extended 
is the region encompassed by the multiple solutions. Then, 
above a given gas pressure, again a unique, but this time 
``cold'' solution, arises. Krolik et al. showed that such 
multiple solutions may lead to the existence of a two-phase 
medium consisting of a hot, dilute gas confining denser, cooler 
clumps, which they identified as the Broad Line Region in AGN.  

To better address the thermal instability issue, these 
authors introduced the ionization parameter $\Xi$ (dimensionless), 
defined as 
\begin{equation}
\Xi = \frac{F_{\rm ion}}{n\,kT\/c} \ ,
\label{eq-field}
\end{equation}

where $F_{\rm ion}$ is the 
incident ionizing flux between 1 and $10^3$ Rydberg, 
$n$ is the total numerical density, and $T$ the gas temperature. 
In a fully ionized gas, the radiation-to-gas-pressure ratio, 
$P_{\rm rad}/P_{\rm gas}$, is about equal to $\Xi/2.3$. The 
so-called S-shape curve is thus, actually, the curve giving 
$T$ versus $\Xi$\/.  Krolik et al. 
showed that the multi-branch region of this 
curve (or, which is equivalent, the multi-solutions 
regime) corresponds to a range of $\Xi$ between 1 and 
10 for a typical AGN spectrum. 

Since then, many studies have been devoted to the S-shape 
curve and to its dependence on the incident ionizing spectrum 
and on the abundances, but always in the {\it optically thin} 
case. The purpose of our study is different from previous works, 
as we are interested in  {\it optically thick}\footnote{Note 
that by ``optically thick", we here mean media optically 
thick to the photoionization continuum; this is the case for 
gas irradiated by a continuum that gets absorbed at 
one or more ionization edges, being therefore altered when 
passing through the medium. In a weakly ionized medium, 
this can occur for a column density as small as 
$10^{18}$\,cm$^{-2}$.}  media such as the 
Warm Absorber (WA) in type 1 AGN, the X-ray line-emitting 
gas in type 2 AGN, or irradiated accretion discs in AGN 
and X-ray binaries. Our studies may be applied to media in   
any pressure equilibrium conditions, e.g., constant gas pressure, 
constant total\footnote{In this paper, the 
total pressure includes the gas and radiation pressure, only; 
however, it would be possible to include other contributions, like 
a turbulent and/or a magnetic pressure component to the total 
pressure.} pressure, or hydrostatic pressure equilibrium. 

\subsection{Optically thick media}
Let us assume a medium in total pressure equilibrium, consisting 
of a slab of gas with a given density at the 
illuminated surface. Under such conditions, 
irradiation by X-rays induces a thermal instability 
beyond a given layer in the slab. The shape of the radiation 
spectrum and the radiation pressure itself, depend on the 
considered position in the gas slab. Near the surface, the radiation 
spectrum is close to the incident spectrum irradiating the slab 
(close, yet different, owing to the additional presence of  
radiation returned by the non-illuminated side of the slab). 

If the radiation-to-gas-pressure ratio is sufficiently large, 
the equilibrium temperature is high, and these layers are hot, 
displaying a unique, ``hot'' solution. Such equilibrium temperature 
stays almost constant in this region. In deeper layers, two 
effects take place. First, the radiation spectrum is harder, 
owing to soft X-ray absorption by the previous layers; it  
contains only hard X-rays and diffuse ultraviolet radiation 
created within the slab. As a consequence, the S-shape 
curve is more pronounced. Second, in the conditions of our 
models (described in Sec.~\ref{models}), the radiation flux 
decreases --- and so does the radiation pressure --- 
entraining the gas to enter the multi-solutions regime; 
the gas has now the choice between two stable solutions, a 
``hot'' and a ``cold'' one. 
Finally, as the radiation penetrates still deeper in the slab, 
it suffers more and more modifications, and we get to a point where 
there is again a unique,  ``cold'' solution. The corresponding 
equilibrium temperature remains almost constant again. 
This phenomenon can occur several times in the slab, 
according to the spectral distribution of the specific 
intensity through the medium.  Each time  the gas-phase 
changes, adjusting to a colder solution,  the temperature 
profile displays a sharp drop by one order of magnitude or 
so. In between phase changes, the temperature remains more 
or less constant.

\subsection{Electronic conductivity and alternative methods}

One of the questions one may ask at this stage is 
``which  solution to choose when there are two or more 
stable ones?'' The answer depends on the previous 
history of the medium. A unique equilibrium state could be  
found if electron conduction was to be included 
(Begelman \& McKee 1990; McKee \& Begelman 1990; 
\agata\ 1999). The well-known numerical 
difficulty linked to the thermal instability 
could be overcomed using an integral formulation of the 
criterium giving the equilibrium solution, instead of 
a differential one (\agata\ \& Czerny 2000).  
However, coupling conductivity effects with a complete 
radiative transfer in photoionization models is a complicated 
task, and has not yet been performed. Attempts to introduce 
electonic conduction are presently being undertaken by 
Chevallier et al. (in preparation). 
 
There are, however, other means to circumvent the problem 
of multiple stable solutions when dealing with photoionization 
codes. Two possibilities are discussed here: the first one was 
used to compute the vertical structure of irradiated accretion 
discs in hydrostatic equilibrium (e.g., Ko \& Kallman 1994). 
This method consists in keeping the gas pressure constant during 
the iterations for convergence of the energy balance equation; 
this procedure is correct, but leads to a 
situation of thermal instability, as described in the previous 
sections. Photoionization codes using 
this computational method (e.g. XSTAR: Kallman \& Krolik 1995; 
Kallman \& Bautista 2001) 
must then choose arbitrarily one of the stable solutions in 
the multi-branch region. 
The second possibility is to keep the density 
constant during the iterations for convergence of 
the energy balance equation, allowing to obtain the 
temperature value in the layer; this computational method provides 
a unique solution, even if it is an approximate one, 
intermediate\footnote{This ``intermediate'' solution 
should not be identified with the unstable solution 
(for which the derivative of ${\cal L}$ is negative);  
it is simply a numerical alternative, intermediate 
between the stable solutions.}  
between the ``cold'' and the ``hot'' solutions present in 
the multi-branch region. This 
approach has been used in studies of the 
vertical structure of accretion discs (Raymond 1993; Shimura, 
Mineshige \& Takahara 1995; Madej \& \agata\ 2000; Kawaguchi, 
Shimura \& Mineshige 2001; Ballantyne et al. 2001). 

Such a computational method is implemented in our 
photoionization code \titan, described in the following 
section. It has been used during the past years, 
with good results; models computed with the 
``intermediate''  temperature solution  have been applied, 
for instance,  to studies of the vertical disc structure 
with hydrostatic equilibrium (\agata\ et al. 2002), and to total 
pressure equilibrium media (\agata\ et al. 2002;  Dumont et al. 
2002; \agata\ et al. 2006; Gon\c{c}alves et al. 2006;  Chevallier 
et al. 2006a).  Since then, we have implemented a new algorithm in 
the \titan\ code; here, we report on this new addition and its 
applications. In summary, the new algorithm allows to choose 
between the ``hot'' or the ``cold'' stable solutions in 
the multi-branch region, and thereof to compute a fully 
consistent model using the chosen solution. For 
the first time, it is thus possible to compare models issued from 
the true stable solutions with the approximate solution model  
and to estimate the degree of uncertainty of such an 
``intermediate'' solution. Our improved version of the \titan\ 
code will allow to  better understand the behavior of both kind 
of computational schemes and to estimate their influence on 
the description of the ionized gas structure and on the 
modelled spectra emitted and absorbed by the medium. 

In Sect.~\ref{titan} we give a brief description of the \titan\ code and 
discuss the different computational methods used in more detail; the  
models are described in Sect.~\ref{models}. Section~\ref{sols} 
contains a comparative study between the ``hot'', ``cold'', 
and ``intermediate'' solutions. In Sect.~\ref{obs} we discuss 
some observational implications, and in 
Sect.~\ref{conclusions} we summarize our conclusions.

\section{Computational issues}\label{titan}

\subsection{The TITAN code}
All our models, summarized in Table~1, were computed using 
the \titan\ code. \titan\ is a photoionization-transfer code 
developed by our team to correctly model optically thick 
(Thomson optical depth up to several tens) ionized media; 
it can equally be applied to thinner media (Thomson depth  
$\sim$ 0.001 to 0.1, e.g., Collin et al. 2004; Gon\c{c}alves et al. 
2006). The code includes all relevant physical processes 
(e.g., photoionization, radiative and dielectronic recombination, 
fluorescence and Auger processes,  collisional ionization, 
radiative and collisional excitation/de-excitation, etc.) 
and all induced processes. It solves the ionization 
equilibrium of all the ion species of each 
element\footnote{Our atomic data include $\sim 10^{3}$ 
lines from ions and atoms of H, He,  C, N, O,  Ne,  Mg, Si, 
S, and Fe.}, the thermal equilibrium, the statistical 
equilibrium of all the levels of each ion, and the transfer 
of both the lines and the continuum. 
It gives as output the ionization, density, and temperature 
structures, as well as the reflected, outward, and transmitted 
spectra. The energy balance is ensured locally with a precision 
of 0.01\%, and globally with a precision of 1\%. 

\begin{table}[b!]
\begin{tabular}{lrrc}
\hline
\hline
Model & $\xi=L/n_{\rm H} R^{2}$ & $N_{\rm H}$~~~ & Thermal instability \\ 
name  & (erg~cm~s$^{-1}$)       & (cm$^{-2}$)    & solution \\
\hline
HI\,1\_C & 10000~~~~~~  & 2\,10$^{23}$   & cold \\ 
HI\,1\_H & 10000~~~~~~  & 2\,10$^{23}$   & hot  \\ 
HI\,1\_I & 10000~~~~~~  & 2\,10$^{23}$   & intermediate \\ 
\hline
MI\,1\_C & 1000~~~~~~   & 2\,10$^{23}$   & cold \\ 
MI\,1\_H & 1000~~~~~~   & 2\,10$^{23}$   & hot  \\ 
MI\,1\_I & 1000~~~~~~   & 2\,10$^{23}$   & intermediate \\ 

MI\,2\_C & 1000~~~~~~   & 2.5\,10$^{23}$ & cold \\ 
MI\,2\_H & 1000~~~~~~   & 2.5\,10$^{23}$ & hot  \\ 
MI\,2\_I & 1000~~~~~~   & 2.5\,10$^{23}$ & intermediate \\ 

MI\,3\_C & 1000~~~~~~   & 3\,10$^{23}$   & cold \\ 
MI\,3\_H & 1000~~~~~~   & 3\,10$^{23}$   & hot  \\ 
MI\,3\_I & 1000~~~~~~   & 3\,10$^{23}$   & intermediate \\ 
\hline
LI\,1\_C & 300~~~~~~    & 5\,10$^{22}$   & cold \\ 
LI\,1\_H & 300~~~~~~    & 5\,10$^{22}$   & hot  \\ 
LI\,1\_I & 300~~~~~~    & 5\,10$^{22}$   & intermediate \\ 

LI\,2\_C & 300~~~~~~    & 8.5\,10$^{22}$   & cold \\ 
LI\,2\_H & 300~~~~~~    & 8.5\,10$^{22}$   & hot  \\ 
LI\,2\_I & 300~~~~~~    & 8.5\,10$^{22}$   & intermediate \\

\hline\hline
\end{tabular}
\caption{Models computed with \titan\ in the framework 
of this study. In the first column, the model name regroups 
information on the ionization (H:~high, M:~medium, L:~low) 
and on the chosen computational solution (C:~cold, H:~hot, 
I:~intermediate). The ionization parameter and column 
density values are given in Cols. 2 and 3, respectively. 
Col. 4 lists the thermal instability solution chosen.
}
\end{table}

\titan\ is based on an idea initially depicted in Collin-Souffrin  
\& Dumont (1986); its conception took several years, the code 
being described for the first time in Dumont, Abrassard \& 
Collin (2000). It has been permanently upgraded since then; some of 
the improvements are described in, e.g.,  Dumont \& Collin (2001), 
Dumont et al. (2002), and Chevallier et al. (2006a). In particular, 
our code uses the Accelerated Lambda Iteration (ALI) method, 
which allows for the exact treatement of the transfer of both 
the continuum and the lines (see Dumont et al. 2003 for a description 
of the ALI method in the modelling of the X-ray spectra of AGN and 
X-ray binaries).  This is a major improvement  over other 
photoionization codes such as Cloudy (Ferland et al.  1998), 
{{\sc XSTAR}} (Kallman \& Krolik 1995; Kallman \& Bautista 2001), or 
{{\sc ION}} (Netzer 1993,\,1996) which use, at least for the lines, an 
integral formalism called  the ``escape probability approximation". 
While ALI  computes very precisely line and continuum fluxes, 
treating them in a consistent way, in approximate methods  
the computation of the absorption and emission lines is uncoupled. 
Furthermore, ALI enables a multi-direction utilisation of the code, 
allowing for any illumination angle (including the  
normal direction) and any direction of the outward and reflected 
emission. This fully operational version of the code offers the 
possibility to model an ionized medium in constant density, 
constant gas pressure, or constant total (i.e. gas plus radiation) 
pressure. It has recently been used to study the soft X-ray 
spectra of AGN (Chevallier et al. 2006a) and applied to model 
real data as, for instance, the Warm Absorber in NGC~3783 with a single 
medium in total pressure equilibrium (Gon\c{c}alves et al. 2006),  
or the spectra of bright ultraluminous X-ray sources 
(Gon\c{c}alves \& Soria 2006).

\begin{figure}[t!]
\begin{center}
\includegraphics[width=7.7cm,angle=-90]{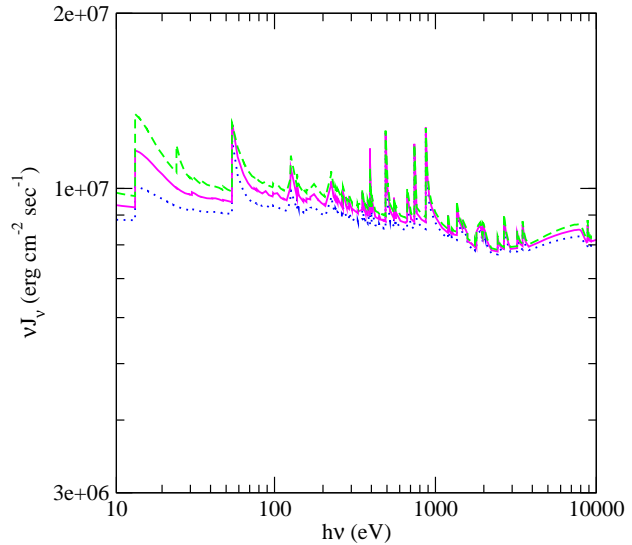}
\caption{This figure displays the spectral distribution of the 
mean continuum intensity, computed at the illuminated surface of 
the slab, for three ``intermediate'' models:   
MI\,1\_I (dotted line), MI\,2\_I (solid line), 
and MI\,3\_I (dashed line). 
For the sake of clarity, the spectral lines have been supressed from 
this figure, where only the continuum is shown; this displays  
important discontinuities, which increase with the slab thickness.}
\label{p8-x3-jcontm1}
\end{center}
\end{figure}

\subsection{Different computational methods used}

We recall that the set of equations describing the physical state of 
the gas (i.e. the local balance between the ionization and 
recombination processes, excitations and de-excitations, as well 
as the local energy balance) is computed at each depth in the 
gas slab. In \titan`s previous computational scheme, the thermal energy 
balance equation was solved for a given density at a given depth 
(that of the previous iteration); this method provides a unique 
solution for the temperature. The model corresponding to this 
scheme will be called ``intermediate'' model hereafter.  

In the new version of \titan, the energy balance 
equation is solved for a given total pressure, assumed 
to be constant along the slab. In some regions of the cloud, 
this computational scheme leads to two stable solutions, 
described by two separate models.  
In one of the models, the ``hot'' solution is systematically 
chosen for all layers, while in the other model the ``cold'' 
solution is systematically preferred. These models will be 
called hereafter ``hot'' model and  ``cold'' model, respectively. 
They actually represent two extreme results corresponding to 
a given gas composition and photoionizing flux, being compatible 
with the two stable solutions. It is important to stress 
that {\it these models differ not only in the layers where 
multiple solutions are possible, but all along the gas slab}, 
this because the entire radiation field will be modified in 
a thick medium. 

\begin{figure*}[t!]
\begin{center}
\includegraphics[width=18cm]{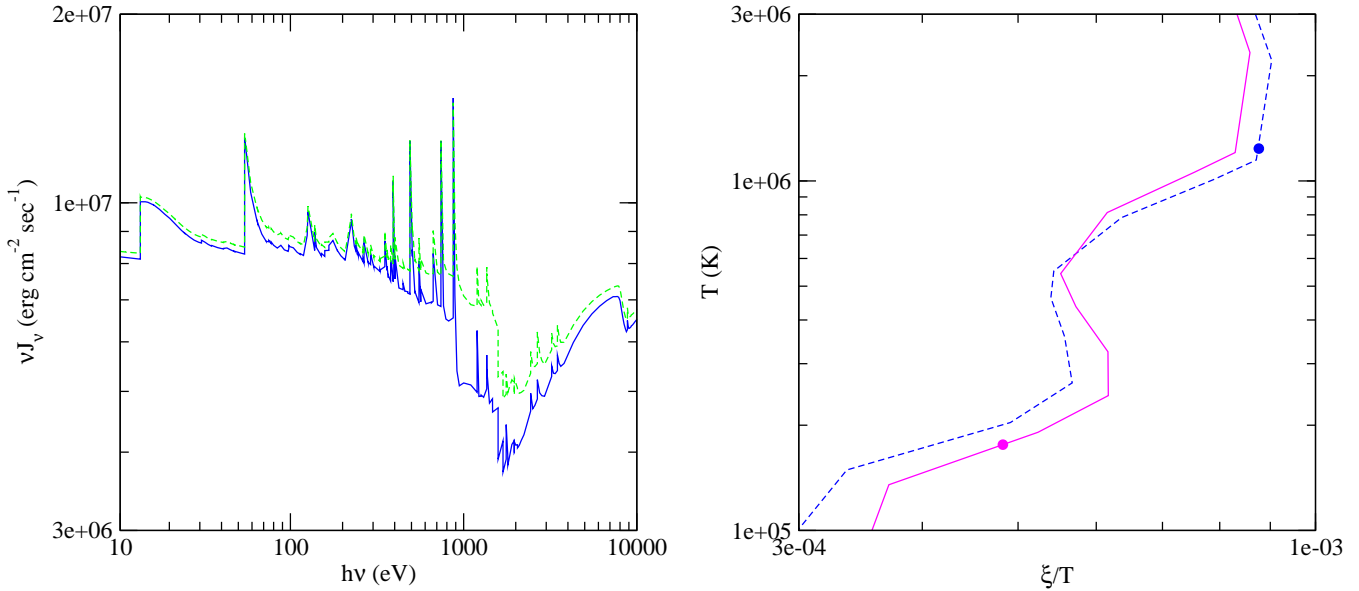}
\caption{This figure displays some results obtained for 
an ``intermediate'' model with $\xi = 1000$ and 
$N_{\rm H} = 2\,10^{23}$ cm$^{-2}$ (MI\,1\_I). On the left-hand panel 
we give the spectral distribution of the mean continuum 
for two layers located at different depths in the gas slab; 
these layers correspond to a column density value of 
1.66\,10$^{23}$ (dashed line) and 1.90\,10$^{23}$ (solid line). 
On the right-hand panel, we display the S-curves giving 
$T$ versus ~$\xi/T$ for the same two layers (the same 
line-codes apply); 
the dots represent the equilibrium temperature found 
for each layer.}
\label{x3c2e23-Jcont-Xicurve}
\end{center}
\end{figure*}

This, and other behaviours of a slab of optically thick gas in 
constant pressure equilibrium, will be discussed in the following  
sections. We have chosen to work with a few models encompassing 
the range of conditions valid for the Warm Absorber in AGN; 
they are  given in Table~1. The same study could, of course, 
be performed for a different set of values; this would not 
change our main results and conclusions.

\begin{figure*}
\begin{center}
\includegraphics[width=18.3cm]{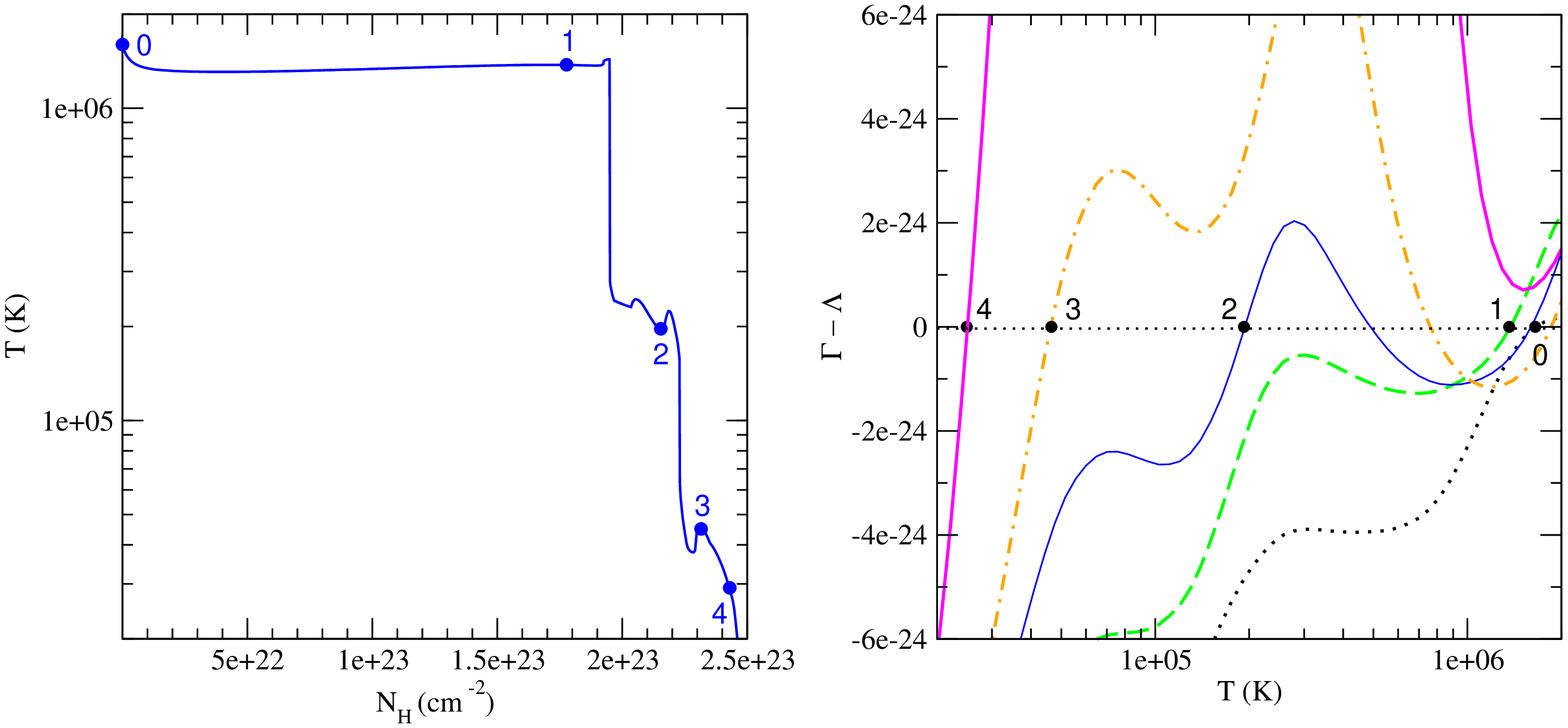}
\caption{On the left-hand panel we give the temperature 
profile along the slab (i.e., versus  
the column density),  for the model MI\,2 ($\xi = 1000$, 
$N_{\rm H} = 2.5\,10^{23}$) computed with the ``cold'' 
solution. The dots represent gas at different depths 
in the slab, for which we have plotted the corresponding 
cooling curves on the right-hand panel. We observe that 
the net cooling function {$\cal{L}$} has one single solution for the 
layers labelled 0, 1 and 4 in the slab, while there are 
three possible solutions for the layers labelled 2 and 3; 
from these, only the ``cold'' solution was selected to 
compute the gas structure and spectra.\vspace{1.5cm}}
\label{p10x3c2_5e23-benef-T}
\end{center}
\end{figure*}

\begin{figure*}
\begin{center}
\includegraphics[width=18.3cm]{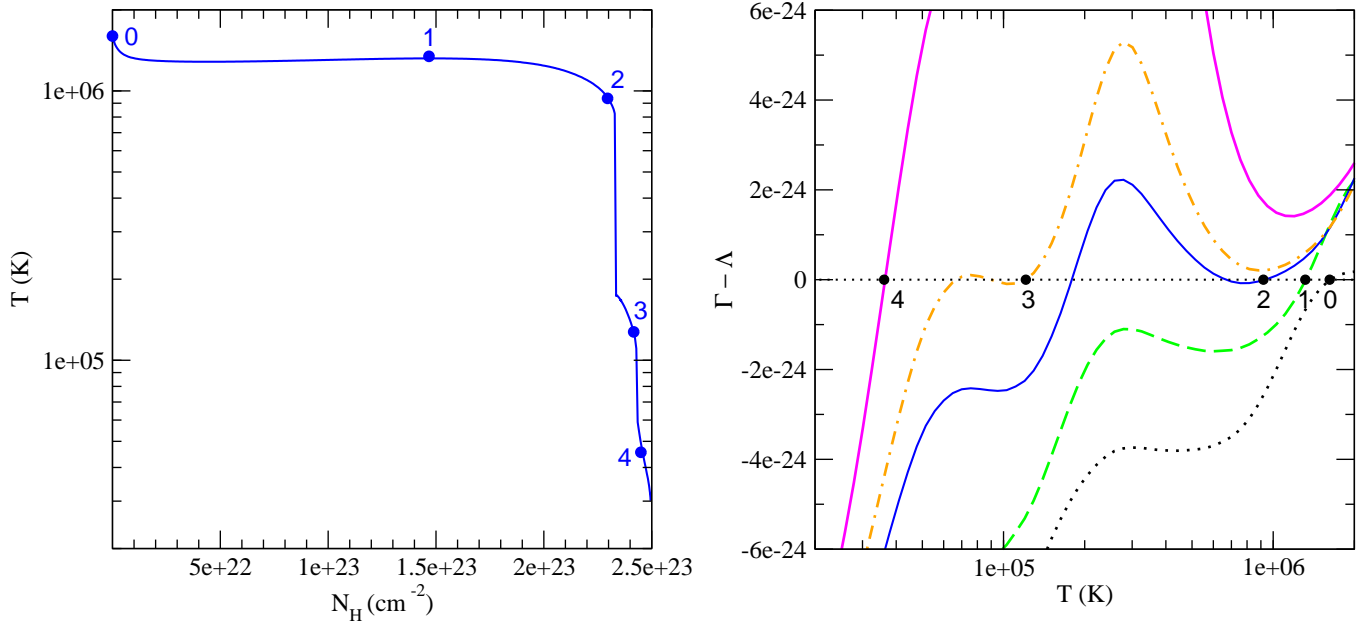}
\caption{Same as in Fig.~3, except the model has 
been computed with the ``hot'' solution. In the right-hand 
panel, 
layers labelled 2 and 3 display three possible solutions, 
from which the ``hot'' one has been systematically choosen; 
the layers labelled 0, 1, and 4 display a single solution.}
\label{p11x3c2_5e23-benef-T}
\end{center}
\end{figure*}

\begin{figure*}
\begin{center}
\includegraphics[width=18cm]{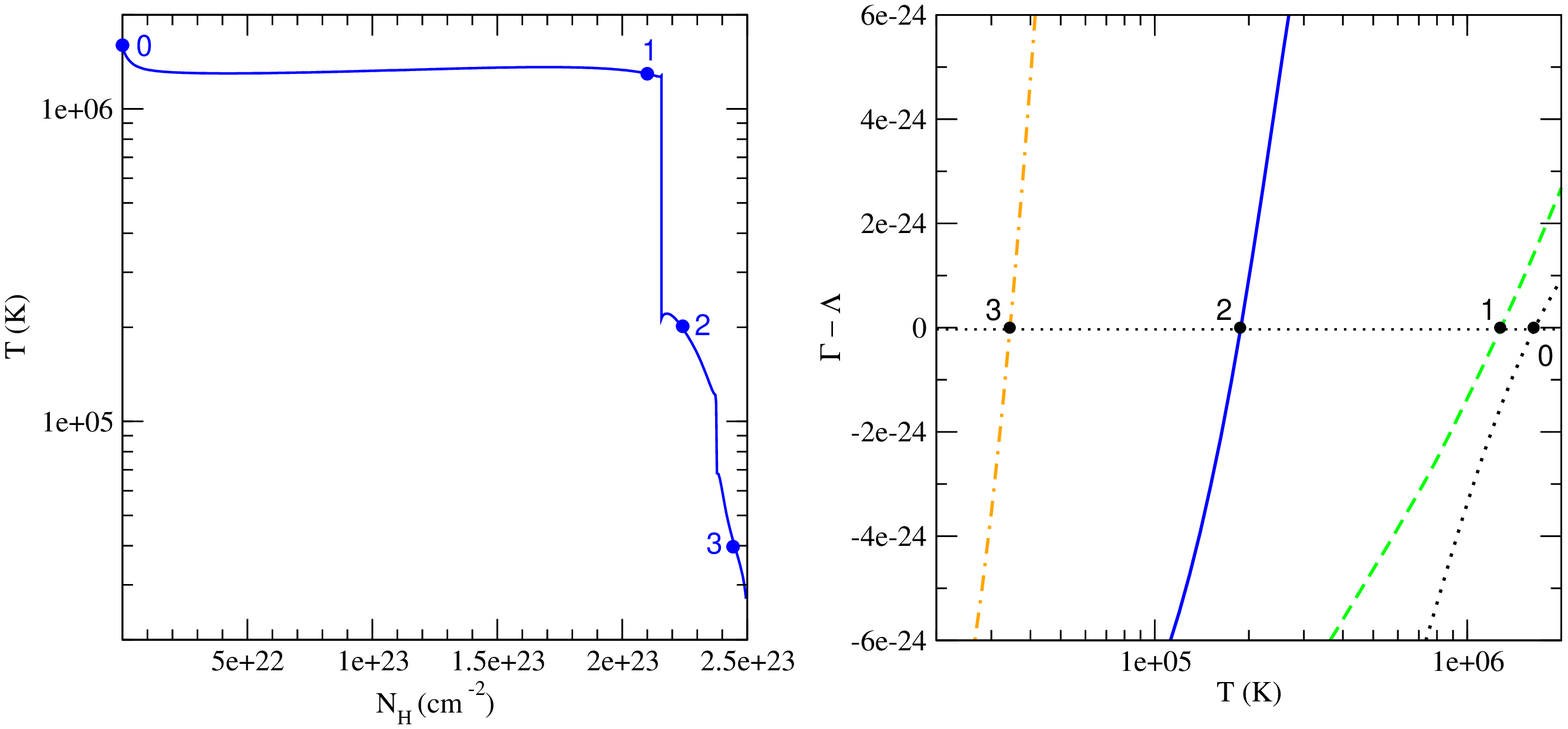}
\caption{Same as in Figs. 3 and 4, except the model 
has been computed with the ``intermediate'' solution. In 
this case, the cooling curves in the right-hand panel 
always provide a unique solution, independently of 
the depth of the layer considered (dots on the 
left-hand panel).}
\label{p8x3c2_5e23-benef-T}
\end{center}
\end{figure*}

\section{The models}\label{models}

All the models presented here were treated in a 1-D plane-parallel 
geometry, as slabs of gas illuminated from one side by a radiation 
field concentrated 
in a very small, pencil-like, shape centered on the normal 
direction. The resultant spectrum, reprocessed by the gas, 
is a combination of ``reflection'' from the illuminated 
side of the medium (not a real reflection, as it includes 
atomic and Compton reprocessing), ``outward emission'' 
(coming from the non-illuminated side of the medium), and 
a ``transmitted'' fraction of the incident ionizing continuum. 
The absorption spectrum (which corresponds to the outward 
emission in the normal direction) depends strongly on 
the column density, since the cool absorbing layers are 
located at the back side of the gas slab.
The relative contribution of each component to the observed 
spectrum depends on parameters such as the size, the density, 
and the geometry of the ionized medium. 

Our models are globally parameterized by the column density 
\NH\ (in units of cm$^{-2}$), the hydrogen number density at 
the illuminated side of the slab \nh\ (in units of cm$^{-3}$), 
the incident flux and SED, and the ionization parameter; this 
parameter is defined differently among authors. In this article, 
we use the $\xi$ form of the ionization parameter, where 
\begin{equation}
\xi = \frac{L}{n_{\rm H} R^{2}}\ \  ~~~~~~{\rm (in\ units\ of\ 
erg~cm~s^{-1})} 
\label{xi} 
\end{equation}
in which  $R$ is the distance between the radiating source 
and the photoionized medium, and $L$ is the source's bolometric 
luminosity (in erg s$^{\rm -1}$); in the following, we integrate 
$L$ over the range 10$-$10$^{5}$~eV, but some authors prefer to give 
$\xi$ with $L$ integrated over the 1$-$1000~Ryd region, as used 
by XSTAR. Appropriate conversions should thus be applied 
if using different forms of the ionization parameter, and  
before any numerical comparison. Note that $\xi$, as well as 
$n_{\rm H}$, are quantities defined at the illuminated surface 
of the slab.  We recall that in constant pressure models the hydrogen 
number density varies along the slab, displaying a profile varying 
inversely with the temperature; also the spectral distribution of 
the ionizing radiation changes across the slab for media in 
pressure equilibrium, as those discussed in this paper.

All our models were computed under total (i.e., gas plus radiation)  
pressure equilibrium; we have neglected all other forms of 
pressure (e.g., turbulent, magnetic, ...) contributing 
to the total pressure. The gas slabs were assumed 
to have cosmic abundances (Allen 1973); the hydrogen numerical 
density at the illuminated surface was set to $10^{7}$~cm$^{-3}$. 
Note that $n_{H}$ is a minor parameter in this study, the overall 
spectrum being only proportional for values varying from 
10$^7$ to 10$^{12}$ cm$^{-3}$ (this is usually the case, provided 
the ionizing spectrum is relativelly flat); only the relative 
intensities of the forbidden and permitted emission lines (e.g., 
from He-like ions like \ion{O}{vii}) are sensitive to changes in 
the density.

We have simplified our models by assuming a null turbulent 
velocity component; micro-turbulence has a small influence 
on the thermal and ionization structure for large column densities,  
where the cooling is dominated by bound-free transitions. 
The thermal and ionization structures are mainly determined by 
the ionization parameter and the shape of the ionizing continuum. 
All the models in this paper assume an incident continuum described by 
a power-law of photon index $\Gamma=2$, covering the 10$-$10$^{5}$~eV 
energy range.

\section{``Hot'', ``cold'', and ``intermediate'' solutions}\label{sols}

In this section, we illustrate the differences between 
the previously discussed computational methods by comparing 
the results obtained with the  ``hot'', ``cold'', and 
``intermediate'' models.

\subsection{Behaviour at the surface of the gas slab}

It is often assumed that the thermal instability 
problem in thick media is exactly the same as in thin 
media. This is not true, for mainly two reasons. 
First, the spectral distribution of 
the mean intensity $J_{\nu}$ at the illuminated surface 
of the slab is different from the incident spectrum, as 
it equally contains a ``returning" radiation component  
emitted by the slab itself. As an illustration,  
Fig.~\ref{p8-x3-jcontm1} shows the spectral distribution 
of the mean continuum intensity at the  slab surface  
for three  ``intermediate'' models with $\xi=1000$ 
and 3 different values of the column density. For the sake of 
clarity, the spectral lines were suppressed from the 
figure, and only the continuum 
is shown. We observe that the spectrum {\it at the surface} 
contains strong discontinuities, whose amplitude increases  
with the slab thickness. More generally, the intensity of 
the whole spectrum increases with the slab thickness (an  
expected behaviour since the radiation emitted by a thicker 
slab should be larger). Second, the spectral distribution 
changes as the radiation progresses inside the medium. 
As a consequence, the shape of the S-curve also changes, 
and instead of traveling along a given S-curve as 
$P_{\rm rad}/P_{\rm gas}$ decreases, the temperature 
follows successively different curves. This behaviour 
is illustrated  by the MI\,1\_I model in 
Fig.~\ref{x3c2e23-Jcont-Xicurve}, which 
shows the spectral distribution of the mean continuum for 
two layers located at different depths in the gas slab 
(left-hand panel) 
and the corresponding curves giving $T$ versus $\xi/T$ 
(right-hand panel), which is the same as representing 
the usual S-shape curve. 
We can see that the S-curves are different for the 
two represented layers and, in particular, 
that the curve corresponding to the deeper layer (solid line) 
displays a larger multi-branch region; this is due to its 
larger absorption trough.  The dots represent 
the equilibrium temperature found for each layer.

\subsection{Comparison between cooling curves}

Figures \ref{p10x3c2_5e23-benef-T}, \ref{p11x3c2_5e23-benef-T}, 
and \ref{p8x3c2_5e23-benef-T} concern a slab of gas with 
total column density equal to $2.5\ 10^{23}$ cm$^{-2}$, 
ionized by an incident continuum with $\xi=1000$ (MI\,2 model). 
These figures display, for respectively the ``cold'', ``hot'', and 
``intermediate'' solution models, the temperature profile versus the 
column density (left-hand panel) and the net cooling function 
(${\cal{L}} = \Lambda - \Gamma$) versus the temperature, at different depths 
in the gas slab (right-hand panel); the various layers represented here 
have been marked and labelled in the left-hand panel. By comparing 
both plots we can verify that $\cal{L}$ cancels for the 
equilibrium temperatures observed in the gas structure.

\begin{figure}[!t]
\begin{center}
\includegraphics[width=7.5cm]{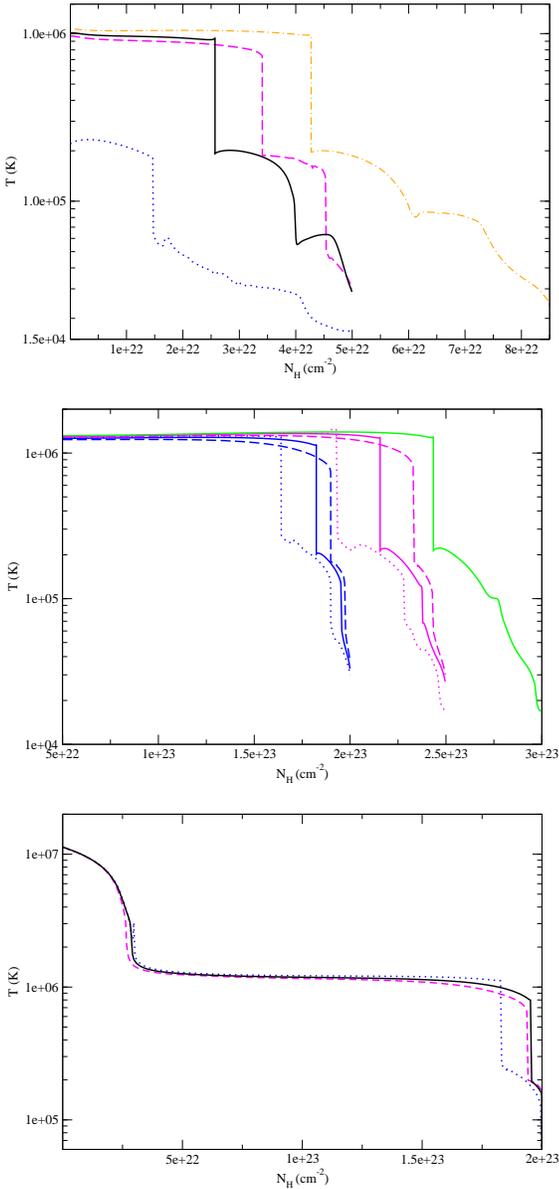}
\caption{Temperature profiles versus the column density for a set 
of models with different $\xi$ and $N_{\rm H}$.  
On the top panel, the LI\,1 models 
(dashed line: ``hot'' solution; dotted line: ``cold'' solution;  
solid line: ``intermediate'' solution); for comparison, we 
also plotted the LI\,2\_I model (dashed-dotted line). On the 
middle panel, several MI models for different values of the 
column density (dashed lines: ``hot'' solution models; dotted lines: 
``cold'' solution models;  solid lines: ``intermediate'' 
solution models). Finally, on the  bottom panel, the HI\,1 models 
(the same line-codes apply). }
\label{T-profile}
\end{center}
\end{figure}

We recall that the cooling curves of the ``hot'' and ``cold'' models 
are determined for a given total pressure, while those of 
the ``intermediate'' models are computed for a given density. 
We can check that the isobaric cooling curves of the ``cold'' 
and ``hot'' models (Figs. \ref{p10x3c2_5e23-benef-T} 
and \ref{p11x3c2_5e23-benef-T}, respectively) display three solutions for 
${\cal{L}} = 0$, while the isodensity cooling curves of the 
``intermediate'' model (Fig. \ref{p8x3c2_5e23-benef-T}) 
always show a unique solution. We stress that it is impossible 
to compare the cooling curves in exactly the same conditions, 
as the models computed with these three solutions differ considerably.  
Indeed, the radiation field is completely different in the three 
models, even at the illuminated surface layer; therefore, 
the isobaric cooling curves differ strongly, even for a layer 
located at a similar depth in the gas slab. 

Let us examine first the cooling curves corresponding 
to the ``cold'' model (Fig.~\ref{p10x3c2_5e23-benef-T}). 
We can see that, at the illuminated surface and at position 1, 
there is a unique solution, while at positions 2 and 3 there 
are three possible solutions, from which the coldest one is chosen; 
near the back surface of the slab, at position 4, the solution is 
again unique. Similarly, 
for the ``hot'' model (Fig.~\ref{p11x3c2_5e23-benef-T}), 
the solution is unique at the surface and at position 1, but 
at positions 2 and 3 the curve has three solutions, from 
which the hotter one is chosen; 
then, at position 4, again the solution is unique.  
However, the multi-branch region is different in the case of the 
``hot'' or the ``cold'' model; this shows how much these two 
models differ.  
Finally, in the case of the 
``intermediate'' model (Fig.~\ref{p8x3c2_5e23-benef-T}), all 
solutions are unique. Given these differences, we would expect 
that the previous computational scheme used by \titan, with its  
isodensity cooling curves, would be completely wrong; this is not 
the case, as we will see later on. Indeed, such a method provides 
temperature structures and spectra which are intermediate between 
those of the ``cold'' and the ``hot'' models.

\begin{figure*}
\begin{center}
\includegraphics[width=18cm]{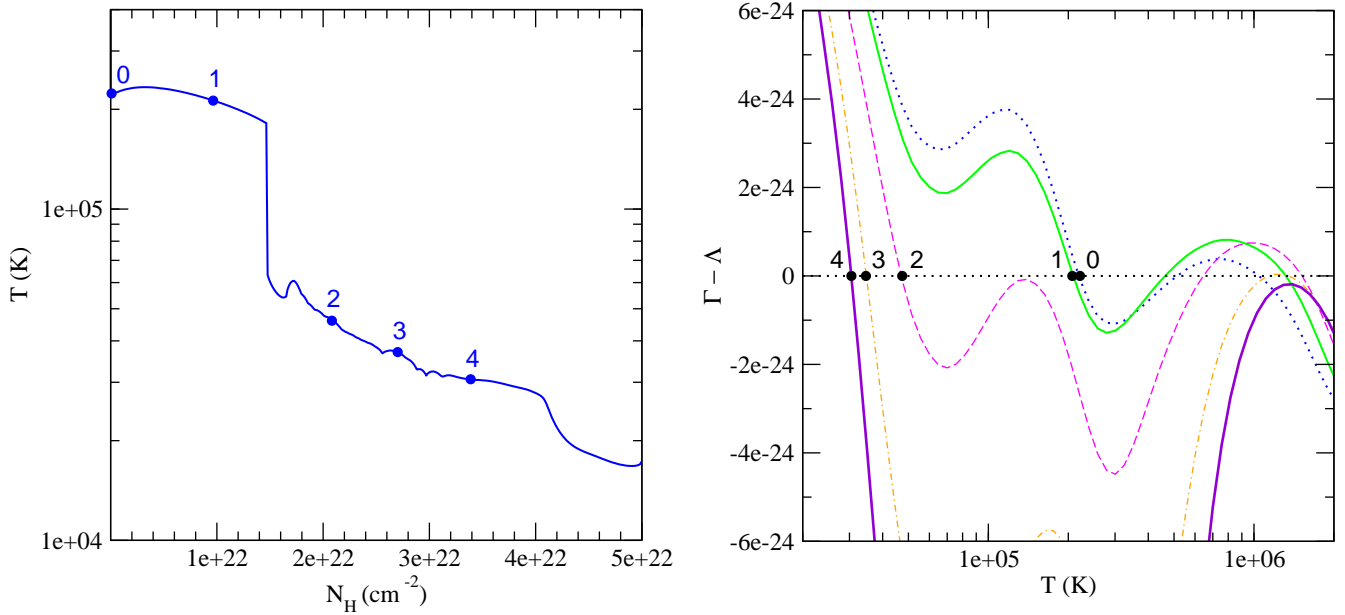}
\caption{Temperature profile versus the 
column density for a model with $\xi = 300$ 
and $N_{\rm H} = 5\,10^{22}$ (left-hand panel). We have marked 
several layers, corresponding to different positions in the 
gas slab, labelled from 0 (at the illuminated surface) to 4  
(deeper in the slab). On the 
right-hand panel, we give the corresponding cooling curves, 
which display one (position 4) or more solutions. 
It is interesting to note that, at such low ionizations, 
the surface layer is already in the multi-solutions regime,  
displaying 3 possible solutions from which the coldest one 
has been chosen.}
\label{p10x300c5e22-benef-T}
\end{center}
\end{figure*}

\begin{figure*}
\begin{center}
\includegraphics[width=18cm]{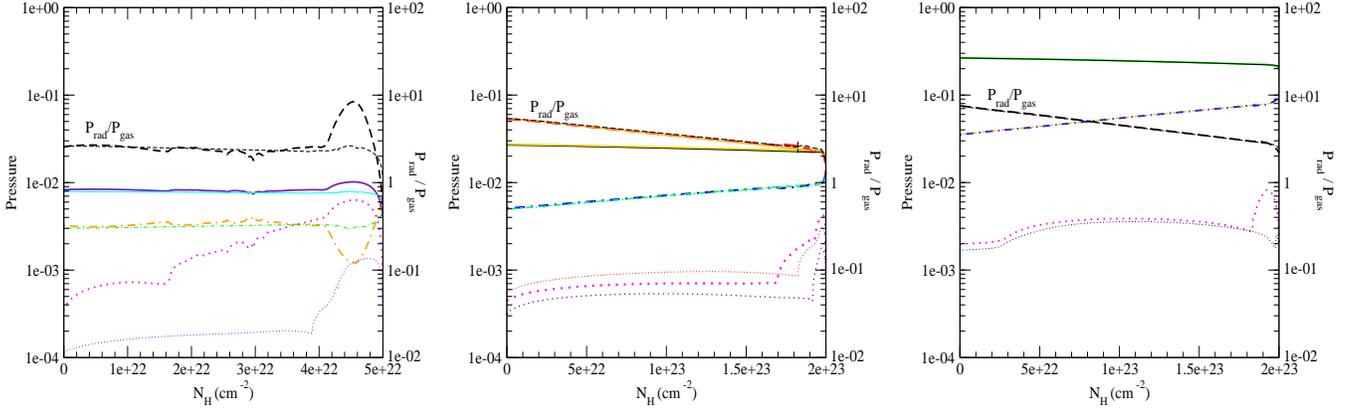}
\caption{This figure shows various pressure quantities computed 
with the ``cold'' (thick lines), ``hot'' (thin lines), 
and ``intermediate''  (thinner lines) solutions for the 
models LI\,1 (left-hand panel), MI\,1 (middle panel), and 
HI\,1 (right-hand panel).  For the sake of clarity, 
the ``intermediate'' solution values are only shown  
for one of the models (MI\,1, middle panel). In all 
the panels, $P_{\rm rad}$ is given in solid lines, 
$P_{\rm gas}$ in dashed-dotted lines, 
$P_{\rm rad}({\rm lines})$ in 
dotted lines, and the $P_{\rm rad}/P_{\rm gas}$ ratio in 
dashed lines; also, the left-hand y-axis 
(10$^{-4}$ $-$ 1) corresponds to the $P_{\rm rad}$, 
$P_{\rm gas}$, and $P_{\rm rad}({\rm lines})$ 
values, while the right-hand y-axis 
(10$^{-2}$ $-$ 10$^{2}$) corresponds to the 
$P_{\rm rad}/P_{\rm gas}$ ratio.} 
\label{pressions}
\end{center}
\end{figure*}

\subsection{Temperature profiles}

Figure \ref{T-profile} displays the temperature profiles 
for the ``cold'', ``hot'', and ``intermediate'' solutions, 
for a set of LI ($\xi = 300$), MI ($\xi=1000$), and HI 
($\xi=10000$) models. The ``intermediate'' solutions 
are represented by solid lines (on the top panel, 
the LI\,2\_I model was represented by a dashed-dotted 
line for the sake of clarity), the ``cold'' solutions  
by dotted lines, and the ``hot'' solutions by dashed lines. 
By analysing this figure, two results appear clearly:

\begin{itemize}

\item First, the differences between the ``hot'', ``cold'', and 
``intermediate'' models increase as $\xi$ decreases; 
\item Second, the temperature always decreases strongly near 
the non-illuminated surface of the slab,  whatever the 
column density.

\end{itemize}

In the following, we will present some clues allowing 
to understand these behaviors.

\subsubsection{Influence of the ionization parameter}

For the higher ionization models ($\xi = 10000$), 
the ``hot'', ``cold'', and ``intermediate'' solutions 
provide  very similar results 
(cf. Fig. \ref{T-profile}, bottom panel). On the contrary, 
for the lower ionization models ($\xi=300$) the results are 
quite different, the first layer at the illuminated side 
of the slab being already in the multi-solutions regime 
(cf. Figure \ref{T-profile}, top panel). In this case, 
since the ``hot'' and ``cold'' solutions correspond to 
very different equilibrium temperatures, there is also a 
large difference on the whole medium structure.  
This is illustrated by Fig.~\ref{p10x300c5e22-benef-T}, 
which is similar to figures 3, 4 and 5, except we now address 
the case of the LI\,1 ``cold'' solution model. The figure shows 
that, already at the illuminated surface of the slab (labelled 0 
on the left-hand panel), the net cooling curve cancels for 
three values of the temperature; this  multi-solutions 
regime lasts until the deeper position labelled 4 on 
the left-hand panel.

\begin{figure}[t!]
\begin{center}
\includegraphics[width=7.7cm,angle=-90]{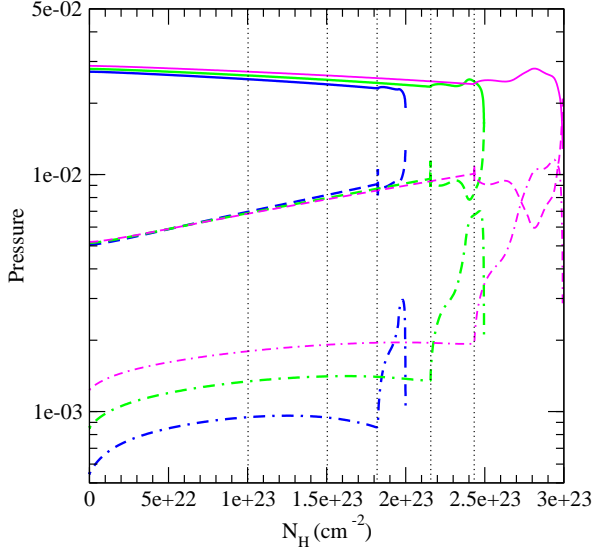}
\caption{Figure showing the variation of 
 $P_{\rm rad}$, $P_{\rm rad}({\rm lines})$, 
and $P_{\rm gas}$ versus the column density, 
for 3 ``intermediate'' models of different 
thickness (MI\,1, MI\,2, and MI\,3). 
$P_{\rm rad}$ is represented by solid lines, 
$P_{\rm gas}$ by dashed lines, and $P_{\rm rad}({\rm lines})$ 
by dashed-dotted lines.  The variation of 
$P_{\rm rad}$ along the gas slab  is very similar 
for the three different thicknesses, until 
the first temperature jump occurs. We also observe that 
near the back side of the gas slab, $P_{\rm rad}$ can 
become very important, owing to the contribution 
from the lines. The thin dotted vertical lines represent 
layers at different depths in the slab, corresponding 
namely to $1\,10^{23}$ cm$^{-2}$, 1.5 $10^{23}$ cm$^{-2}$, 
and to the first temperature jump in each model 
(1.82\,$10^{23}$, 2.16\,$10^{23}$, and 2.43\,$10^{23}$ 
cm$^{-2}$ for the MI\,1, MI\,2, and MI\,3 model, 
respectively). }
\label{p8x3-compar-pre}
\end{center}
\end{figure}

It is possible to address this situation and try to 
find an adequate explanation for the behaviour of 
the LI\,1 model by looking into what's happening 
with the $\Xi$ ionization parameter (or, which is similar, 
studying the $P_{\rm rad}/P_{\rm gas}$ ratio). 
Since the multi-solutions regime occurs for a given range in $\Xi$,  
one expects that the associated region  
in the slab would correspond to layers where 
$P_{\rm rad}/P_{\rm gas}$ is of the order of a few units.  
This is indeed the case, as shown in Fig.~\ref{pressions}. 
This figure displays the variation of $P_{\rm rad}$, $P_{\rm gas}$, 
and $P_{\rm rad}/P_{\rm gas}$ along the slab, for the ``hot'' 
and ``cold'' solutions of a set of LI, MI, and HI models.  
The radiation pressure due to the lines, $P_{\rm rad}({\rm lines})$, 
is also shown, as this component can be quite important near the back 
side of the slab; note that $P_{\rm rad}$, also represented in 
the figure, already includes the contribution from 
$P_{\rm rad}({\rm lines})$.  In this figure, 
the ``cold'', ``hot'', and ``intermediate'' models are 
represented by thick (green, black, violet), 
thin (blue, orange, magenta), and very thin lines 
(cyan, yellow, red), respectively.  
$P_{\rm rad}$ is given in solid lines, $P_{\rm gas}$ in 
dashed-dotted lines, $P_{\rm rad}({\rm lines})$ in 
dotted lines, and the $P_{\rm rad}/P_{\rm gas}$ ratio 
in dashed lines. The left-hand y-axis 
(10$^{-4}$ $-$ 1) corresponds to the $P_{\rm rad}$, 
$P_{\rm gas}$, and $P_{\rm rad}({\rm lines})$ 
values, while the right-hand y-axis (10$^{-2}$ $-$ 10$^{2}$) 
corresponds to the $P_{\rm rad}/P_{\rm gas}$ ratio. 
 For the sake of clarity, the 
``intermediate'' solution  values (thinner lines: 
cyan, yellow, red) are only shown for one of the 
models (MI\,1, in the middle panel).

\begin{figure}[t!]
\begin{center}
\includegraphics[width=7.7cm,angle=-90]{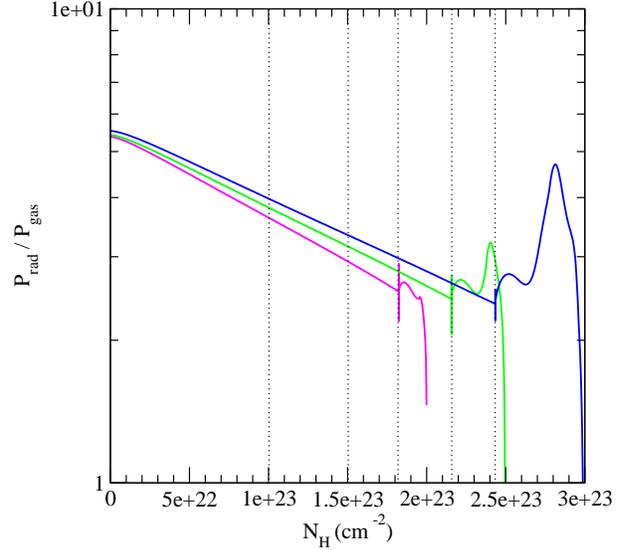}
\caption{Figure showing the $P_{\rm rad}/P_{\rm gas}$ ratio 
versus the column density, for the 3 ``intermediate'' 
models discussed in the previous figure. This  
ratio decreases as we penetrate in the gas slab until it reaches 
a value $\sim$2.5, coinciding with a temperature jump. The figure 
shows that this ratio decreases more rapidly for the thinner, 
than for the thicker  gas slab. 
The thin dotted vertical lines represent layers at different 
depths in the slab (same as in Fig.~9).}
\label{p8x3-compar-pradspgaz}
\end{center}
\end{figure}

Observing Fig.~\ref{pressions}, we first notice that, 
in all cases, the pressure is entirely dominated by  
$P_{\rm rad}$, and that $P_{\rm rad}/P_{\rm gas}$ at the 
illuminated surface increases 
with $\xi$, as expected. The $P_{\rm rad}/P_{\rm gas}$ ratio 
decreases accross the slab from the illuminated surface to 
the back side as the radiation is absorbed, the slab 
entering the multi-solutions regime when $P_{\rm rad}/P_{\rm gas}$ 
is of the order of 2.5; this is true whatever the pressure ratio 
at the illuminated side. It is easy to see that 
such a value was promply attained in the case of the LI\,1 
model discussed before; as a consequence, we were able to 
observe a  ``cold'' and a ``hot'' solution already at the 
surface of the slab. 

\begin{figure*}[t!]
\begin{center}
\includegraphics[width=18cm]{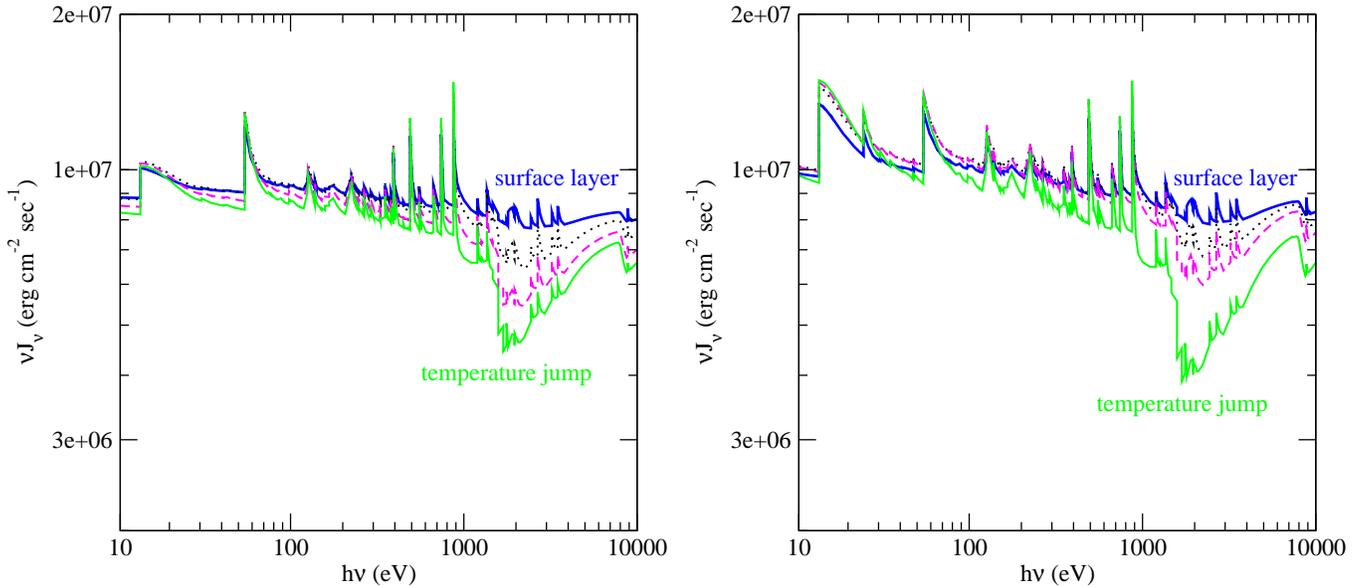}
\caption{Figure showing the medium intensity spectral 
distribution for two MI ``intermediate'' models with different column 
densities: on the left-hand panel, the thiner (shorter) 
model with  $N_{\rm H} = 2\,10^{23}$; on the right-hand panel, 
the thicker (longer) model with $N_{\rm H} = 3\,10^{23}$. 
Both panels display the spectral distribution corresponding 
to 4 layers, from the illuminated surface of the slab (top 
curves) to the layer closer to the temperature jump (bottom curves).}
\label{x3c2-3e23-Jcont-diff-couches}
\end{center}
\end{figure*}

It is also interesting to note that, in spite of a big 
difference in temperature, the gas pressure at the surface is 
almost the same for the ``cold'' and ``hot'' models. This 
behaviour is actually due to an automatic adjustment of the density, 
which is equal to $5\,10^7$ cm$^{-3}$ in the first layer of 
the ``cold'' model, instead of the smaller, initial value of 
$10^7$ cm$^{-3}$. The figure also shows that, for all values 
of the ionization parameter, the pressures in the three models 
(``cold'', ``hot'', and ``intermediate'') are quite similar, 
except for $P_{\rm rad}({\rm lines})$ near the back side of the slab. 
Indeed, the contribution from the lines is always less important in 
a hot medium than in a cold medium, which can emit intense 
ultraviolet lines. Therefore, the smaller the ionization 
parameter, the more important are the lines. It is thus 
easy to understand how $P_{\rm rad}({\rm lines})$ can 
even dominate the radiation pressure near the back side of the 
the LI\,1\_C model (cf. Fig.~\ref{pressions}, 
left-hand panel). 

Note finally that the sharp temperature drops, observed in the 
left-hand panel of figures 3, 4, 5 and 7, do not occur exactly 
at the same value of $P_{\rm rad}/P_{\rm gas}$ for all the models; 
this is due to the shape of the S-curve, and to the different  
spectral distributions inside the slab, which depend not only 
on the type of model (``cold'' or ``hot''), but also on the 
ionization parameter and on the total tickness of the medium.

\subsubsection{Influence of the medium thickness}

As already noted by \agata\ et al. (2006) and Chevallier 
et al. (2006a), the thickness of a pressure equilibrium 
medium cannot exceed a maximum value  for a given ionization 
parameter. This is explained by the fact that, when the 
temperature reaches low values (of the order of 10$^4$~K), 
the radiation pressure becomes dominated by the spectral 
lines, and therefore the gas enters in a new multi-solutions regime. 
When the temperature drops to the cold values corresponding 
to a molecular gas, we reach a limit and are unable to 
provide a solution, as our code cannot handle molecular gas.  

If the maximum thickness of a medium in pressure equilibrium 
is a known issue, there are two other puzzling problems 
related to the thickness of the slab. 
The first is related 
to the fact that the temperature always decreases rapidly 
near the non-illuminated side of the gas slab. 
The second concerns the reason why the slab always enters 
a multi-solutions regime just before the slab ends, even 
when the imposed column density is much smaller than the 
possible maximum value. 

We are going to address these issues using 
the ``intermediate'' solution models as an illustration.  
Let us consider a set of models with 
$\xi =1000$. Figure~\ref{p8x3-compar-pre} displays 
the variation of $P_{\rm rad}$, $P_{\rm rad}({\rm lines})$, 
and $P_{\rm gas}$ along the slab for three values of the column 
density. $P_{\rm rad}$ is represented by  solid lines, 
$P_{\rm gas}$ by dashed lines, and $P_{\rm rad}({\rm lines})$ 
by  dashed-dotted lines.  We can see that the dependence of 
$P_{\rm rad}$  --- and therefore also of 
$P_{\rm gas}$, since the sum of $P_{\rm rad}$ and 
$P_{\rm gas}$ is constant ---  on the gas position inside the slab  
is almost the same for the three different thicknesses until 
the first temperature jump occurs. For the sake of simplicity, 
we do not show here $P_{\rm rad}/P_{\rm gas}$, but it 
is clear that this ratio displays always approximately 
the same value ($\sim$\,2.5) at the temperature jump 
(the position of the temperature jump can be seen in 
Fig.~\ref{T-profile}, middle panel). Such temperature changes 
are accompanied by a strong increase in $P_{\rm rad}({\rm lines})$. 
The thicker the slab, the larger is the cold zone, and 
the stronger the increase in $P_{\rm rad}({\rm lines})$. 
One can even observe an inversion of the gas pressure variation 
near the {\it back surface of the slab}, for the thickest 
case (MI\,3 model). 
Very close to the back surface,  
the layers are optically thin for the continuum and the 
lines, which can thus escape from the medium; this induces a rapid 
decrease in the radiation pressure, as observed in the figure; 
it also leads to a strong cooling, causing a rapid decrease 
of the temperature near the back surface of the slab. 
These coupled phenomena give an answer to the first problem.

\begin{figure*}
\begin{center}
\includegraphics[width=18.2cm]{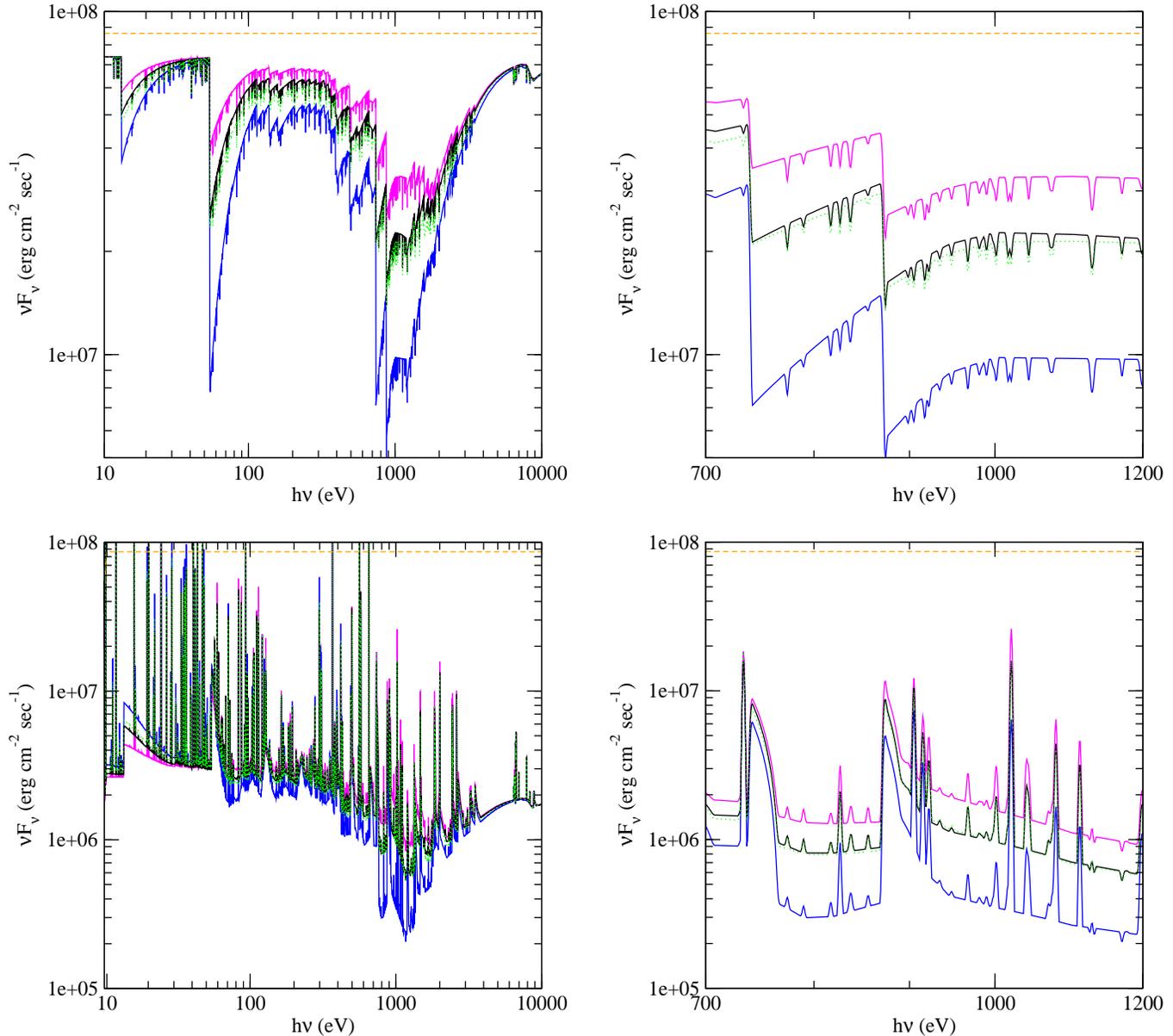}
\caption{This figure gives the pure absorption 
(top panels) and outward emission spectra (bottom panels) 
for the MI\,1 model. The ``cold'' 
solution model is represented by the lower blue lines, the 
``hot'' model by  upper magenta lines and the ``intermediate'' 
model by the middle black lines; the spectra corresponding to 
the half-sum of the ``hot'' and ``cold'' models are 
also plotted for comparison (green dotted lines). The orange 
dashed line corresponds to the incident ionizing continuum.  
The spectra are represented with a resolution of 300; the 
right-hand panel foccuses on the 700$-$1200~eV energy range 
for details, while the left-hand panel shows a larger band  
covering the 10\,$-$\,10$^{4}$~eV range. Notice the 
excellent agreement between the ``intermediate'' and 
half-sum spectra over the entire energy range.}
\label{p8-10-11x3c2e23-spe}
\end{center}
\end{figure*}

The second issue is more subtile.  From Fig.~\ref{p8x3-compar-pre} 
we know that $P_{\rm rad}$ decreases inside the slab 
(except near the back side, owing to the lines, as explained). 
Since $P_{\rm rad}$ is slightly larger in thicker slabs, 
the traversed length before reaching the ratio 
$P_{\rm rad}/P_{\rm gas}$ required for the temperature jump,  
is larger for thicker slabs. However, this is not 
sufficient to explain why the temperature transition is 
located closer to the illuminated surface in thiner slabs, or 
the difference between the transition positions would not 
be so large in media with different thicknesses. Additional 
explanation can be found in Fig.~\ref{p8x3-compar-pradspgaz}, 
which compares the $P_{\rm rad}/P_{\rm gas}$ ratio versus the 
position in the slab (given in terms of column density) for 
the same three slab thicknesses discussed in the previous 
figure; we can see that this ratio decreases more rapidly 
for the thinner (i.e. shorter), than for the thicker (longer) 
gas slab. 

\begin{figure*}
\begin{center}
\includegraphics[width=18.2cm]{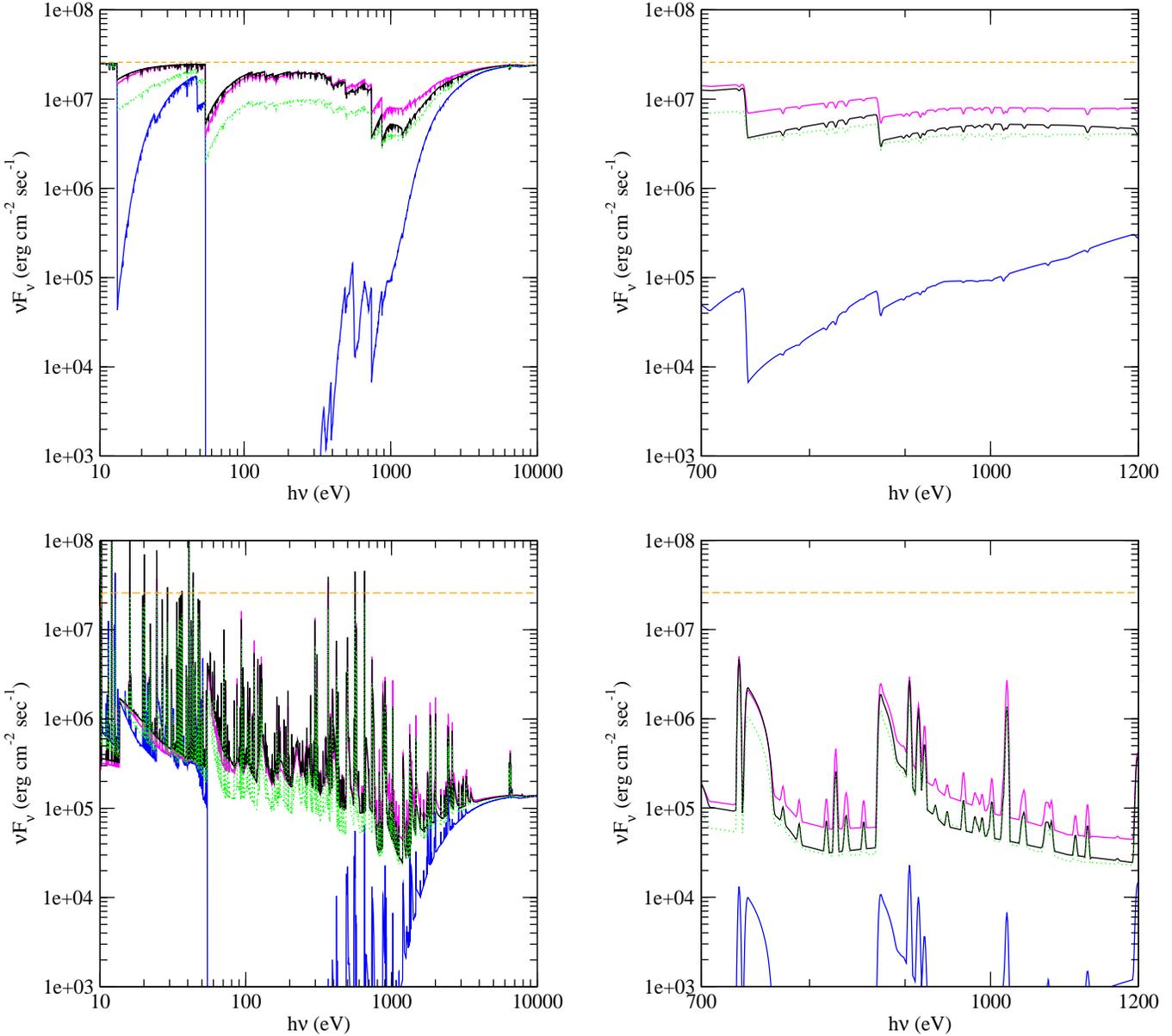}
\caption{Same as in the previous figure, only for the 
lower ionization model LI\,1. In this case, 
the agreement between the ``intermediate'' 
and half-sum spectra is not as good as for the MI\,1 model. The 
differences are less important for the outward emission in the 
X-ray range.  }
\label{p8-10-11x300c5e22-spe}
\end{center}
\end{figure*}

The explanation for this behaviour can be found in the 
comparison between the spectral distribution of the first 
layer, and of a layer located close to the place where 
the jump in temperature occurs. 
Figure~\ref{x3c2-3e23-Jcont-diff-couches} displays 
the spectral distribution of the mean intensity 
(again, the lines were suppessed for clarity), 
for two MI, ``intermediate'' 
models with total column densities 
2\,10$^{23}$ cm$^{-2}$ (left-hand panel) and 
3\,10$^{23}$ cm$^{-2}$ (righ-hand panel), at 
different positions in the slab (from top to bottom: 
at the illuminated surface, at 
$1\,10^{23}$ cm$^{-2}$, $1.5\,10^{23}$ cm$^{-2}$, and at 
the first temperature jump --- $1.82\,10^{23}$ for 
MI\,1 and  $2.43\,10^{23}$ for MI\,3); 
these layers are identified in Fig.~\ref{p8x3-compar-pradspgaz} by 
thin vertical dotted lines. We can observe that the spectral 
distribution of the continuum varies differently in the 
two media: at the same location in the slab, the spectrum 
is less absorbed in the thicker slab (righ-hand panel), 
owing to the more important re-emission component. 
In other words, the spectral distribution of the thiner, 
shorter slab (left-hand panel) has a stronger trough, 
and reaches more rapidly the multi-solutions regime, 
as explained above (cf.~Fig.~\ref{x3c2e23-Jcont-Xicurve}).

\section{Observational implications}\label{obs}
 
\subsection{Outward emitted and absorbed spectra} 

Unfortunately, it is impossible to know what solution the 
plasma will adopt when attaining the multi-solutions regime. 
For instance, it can  oscillate between the hot and cold  
solutions, it can fragment into hot and cold clumps 
which will coexist together, or it can take the form of 
a hot, dilute medium confining cold, denser clumps 
(the hot and cold media sharing the same pressure). 
We recall that the ``cold'' and ``hot'' models computed 
by \titan\ can be interpreted as ``extreme'' models, as 
they adopt respectively the ``cold'' or the ``hot'' solutions  
{\it all along the slab}.  The spectra emitted or absorbed 
by a given ionized medium, consisting of a mixture of gas 
in the hot and cold phases,  should thus be intermediate 
between those resulting from the pure ``cold'' and ``hot'' 
models. The differences between the emitted and 
absorbed spectra obtained with the stable solutions can 
thus provide an indication for the maximum  ``error bars'' 
associated to the spectra computed with the ``intermediate'' 
solution.

As an example, Figs.~\ref{p8-10-11x3c2e23-spe} and 
\ref{p8-10-11x300c5e22-spe} give the pure absorption (top  
panels) and outward emission (bottom panels) spectra 
resulting from the ``cold'', ``hot'', and 
``intermediate'' (black, middle curves) 
solutions for the MI\,2 and LI\,1 models, respectively   
(see Fig.~\ref{T-profile} for the corresponding temperature 
profiles). The spectra are represented with a resolution of 300 
and have been displayed in two energy ranges: 10$-$10$^{4}$~eV 
(left-hand panels), with a zoom in the 700$-$1200~eV region 
(right-hand panel) to more clearly observe the differences. 
The lower solid lines (blue) correspond to the 
``cold'' models, the middle lines (black) to the ``intermediate'' 
models, and the upper lines (magenta) to the ``hot'' models. 
The incident ionizing continuum (orange dashed lines) and 
the spectrum corresponding to the  half-sum of the 
``cold'' and ``hot'' models (green dotted lines) are 
also plotted, for comparison. 
The outward emission spectrum has been computed for an opening angle 
of 45 degrees which, according to the unified Scheme of AGN 
(Antonucci \& Miller 1985), is in agreement with the expected 
geometry for the Warm Absorber in Seyfert~1 nuclei, or the 
emissive region in Seyfert 2s. 

Investigating the MI\,2 spectra in Fig.~\ref{p8-10-11x3c2e23-spe}, 
we can see that the ``intermediate'' model spectra (black solid 
lines, middle) is extraordinary close to the half-sum (green 
dotted lines) of the ``cold'' and ``hot'' spectra, though these 
ones are quite different. The differences amount at most to 
1 or 2\%\ for the outward emitted spectrum, and to 20\%\ 
for the absorption spectrum. The agreement is not as good 
for the second case (LI\,1 model, 
Fig.~\ref{p8-10-11x300c5e22-spe}), where the differences 
between the ``intermediate'' model and the half-sum 
of the ``cold'' and ``hot'' models amount typically to a 
factor 2 for the absorption spectrum. The agreement is 
much better for the outward emission in the X-ray range. 
This is due to the strong imprint of the cold layers near 
the back surface of the ``cold'' model, which absorb almost 
completely the UV-X spectrum. On the contrary, the X-ray 
outward emission, being due mainly to the deep layers, 
is not very different for the ``hot'' and the ``cold'' models. 

We observe that the  lines have similar intensities 
in the ``cold'' and ``hot'' spectra, and also in the ``intermediate'' 
and half-sum spectra; this happens in spite of important 
differences in the continuum intensity, in particular for 
the ionization edges. Such a behaviour is due to the fact that  
the main lines are saturated resonance transitions, while the 
ionization edges are optically thiner, and therefore almost 
proportional to the thickness of the region containing  
the emitting ion (apart from the diffuse emission contribution).

\subsection{Ionization states} 
  
\begin{figure*}
\begin{center}
\includegraphics[width=18.2cm]{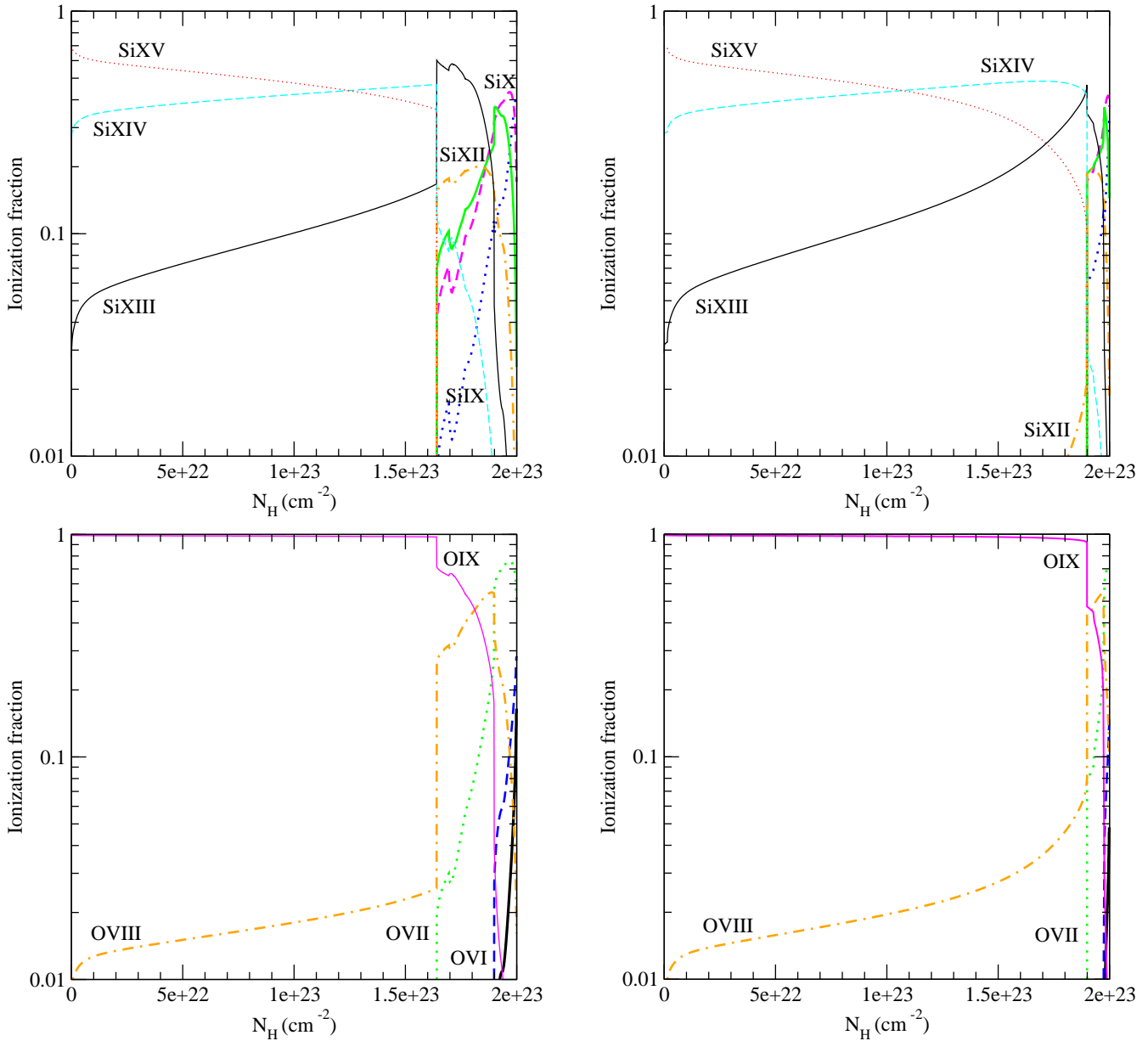}
\caption{Figure showing the fractional 
ionization of the Si ions (top panels) and 
O ions (bottom panels) for the same models as 
in  Fig.~\ref{p8-10-11x3c2e23-spe}. The ``cold'' 
and ``hot'' models are displayed in the left-hand 
panels and right-hand panels, respectively. We 
observe that the regions producing the various 
ions have important size differences in the case 
of a ``cold'', or of a ``hot'' model. }
\label{p10-p11x3c2e23-ion}
\end{center}
\end{figure*}

Differences in the ionization state for ``hot'' and ``cold'' 
models are illustrated by 
Fig.~\ref{p10-p11x3c2e23-ion}, which displays the fractional 
ionization of two important elements --- Si (top panels) and 
O (bottom panels) --- for the ``cold'' (left-hand panels) and 
``hot'' (right-hand panels) models already 
depicted in  Fig.~\ref{p8-10-11x3c2e23-spe}. 
This figure clearly shows why the intensities of the 
ionization edges are different in the ``cold'' and ``hot'' 
models. We can see that the length 
of the \ion{O}{vii} and \ion{O}{viii} regions   
is about three times larger in the ``cold'' model than 
it is in the ``hot'' model, these differences being 
somewhat larger than the discrepancy in the 
corresponding spectra.

\subsection{Variability}

In the multi-solutions regime, the medium should be made 
of a mixture of gas in the hot and cold phases, and the relative 
proportion of those phases could be varying with time. 
The whole medium should thus oscillate between physical 
states located  somewhere in the parameters' region covered 
by the ``hot'' and ``cold'' models. The variation timescale 
in such a medium is basically the thermal time in the 
multi-branch region. However, if the medium is subject to 
a change in temperature, this will not necessarily be immediately  
followed by a change in ionization inducing modifications 
in the main absorbing or emitting ions. It all depends on 
the relative values of the thermal time  $t_{\rm th}$, the 
ionization time $t_{\rm ion}$ and the recombination time  
$t_{\rm rec}$.  One thus expects two types of variability: 
a relativelly weak variability related only to a temperature 
variation, and a stronger variability related to a change of 
ionization state. If $t_{\rm ion}$  or  $t_{\rm rec}$ 
are smaller than $t_{\rm th}$, strong variations in the  
emitted/absorbed spectrum should be observed during the 
thermal time-scale, with the main emission/absorption lines 
and ionization edges being replaced by lines and edges of 
other ions. If $t_{\rm ion}$  and  $t_{\rm rec}$ are 
larger than $t_{\rm th}$, the spectrum should display 
smaller changes, like a simple variation in the 
relative intensity of the spectral features. 
	
Let us compare these three timescales. The thermal time 
$t_{\rm th}$ is of the order of $kT/n{\Lambda}$. 
In the multi-solutions regime, the cooling and the heating  
are dominated by atomic processes, and {$\Lambda$} is of 
the order of 10$^{-23}$ erg\,cm$^{3}$\,s$^{-1}$, 
so $t_{\rm th}$ writes:

\begin{equation}
t_{\rm th} \sim {T_5\over n_{12} \, {\Lambda_{23}}} \ \ \ \ {\rm (in~units~of~s)},
\label{ttherm} 
\end{equation}	

where $n_{12}$ is the density expressed in 10$^{12}$ cm$^{-3}$, 
${\Lambda}_{23}$ is the cooling function in 
10$^{-23}$ erg\,cm$^{3}$\,s$^{-1}$, and $T_5$ is 
the temperature in 10$^5$~K. 

For a given ion, $t_{\rm rec}$ is equal to 
$ t_{\rm rec} = 1/( n_{\rm e}\, \alpha_{\rm ion})$
where $n_{\rm e}$ is the electron numerical density 
(roughly equal to the total density $n$), and 
$\alpha_{\rm ion}$ is the recombination coefficient 
of the given ion, including dielectronic recombinations. 
This coefficient is almost independent from the 
radiation flux, with a typical value  of 10$^{-11}$ cm$^3$\,s$^{-1}$ 
at the temperature of the multi-branch region for a heavy,  
highly ionized element, so $t_{\rm rec}$ writes:

\begin{equation}
t_{\rm rec} \sim {0.1\over n_{12} \, \alpha_{11}}\ \ \ \ {\rm
(in~units~of~s) },
\label{trec} 
\end{equation}	  

where the recombination coefficient $\alpha_{11}$ is 
expressed in 10$^{-11}$ cm$^3$\,s$^{-1}$. Thus, 
the recombination time $t_{\rm rec}$ of a heavy element in 
a highly ionized state is smaller than, or of the order of, 
the thermal time $t_{\rm th}$. One deduces that a thermal 
variation induces almost immediately a change in the ionization state.

Ionization equilibrium is reached after a time equal to 
the maximum time-scales given by $t_{\rm ion}$ and $t_{\rm rec}$. 
The ionization time $t_{\rm ion}$ is defined as: 
\begin{equation}
t_{\rm ion} =\left( \int_{\nu({\rm ion})}^{\nu({\rm max})}
{4\pi \, {J_{\nu}\over h\nu} \, \sigma_{\nu} \, d\nu}\right)^{-1},
\label{tion1} 
\end{equation}	  

where $\sigma_{\nu}$ is the frequency-dependent photoionization 
cross-section of the given ion. As we have seen, $J_{\nu}$ varies 
across the slab,  but we can make a very rough estimation of 
$ t_{\rm ion} $ by assuming that $J_{\nu}$ is of the order of 
the incident flux, $F_{\nu}/2\pi$. This could lead to a large 
overestimation of the ionization rate, and therefore to an 
underestimation of the ionization time, as $J_{\nu}$ is actually 
smaller than $F_{\nu}/2\pi$ in the deep layers, at the back of the 
slab (cf. the outward emitted spectrum in Figs.  
\ref{p8-10-11x3c2e23-spe} and \ref{p8-10-11x300c5e22-spe}). 
With our power law incident spectrum described by 
$F_{\nu} \propto \nu^{-1}$ in the 10$-$10$^{-5}$~eV range, 
and assuming that the cross-section 
is proportional to $\nu^{-3}$, one finds\footnote{To derive 
this equation, we have made use of the definition 
of $\xi$ given in Eq.~\ref{xi}. }:

\begin{equation}
t_{\rm ion} \sim {3\ 10^{-5} \over \xi_3 \, n_{12} \, \sigma_{19}}\ \ \ \ 
{\rm (in~units~of~s)},
\label{tion2} 
\end{equation}	  

where $\xi_3$ is the ionization parameter $\xi$ expressed 
in 10$^3$~erg\,cm\,s$^{-1}$ and $\sigma_{19}$ is the cross-section 
at the ionization edge of the given ion expressed in 
10$^{-19}$~cm$^{-2}$. 

In conclusion, even if it may be underestimated 
by two orders of magnitude, $ t_{\rm ion}$ 
is smaller than $ t_{\rm rec}$, and both 
time-scales are smaller than  $t_{\rm th}$; therefore, if 
the temperature changes locally in the multi-branch region, 
a new ionization equilibrium settles in immediately.

The problem is more complex in practice. Indeed, in order 
to observe variations in the spectrum, the change of 
temperature should affect a large proportion of the 
multi-branch region.  The mass fractions of the hot and cold  
phases are most probably of the same order, so this zone 
should be made of cold, dense clumps embedded in a hotter and more 
dilute medium. In this case, the evaporation time of the clumps 
should govern the structural changes. This time-scale is of 
the order of the cold phase dynamical time, if evaporation is 
not saturated. 
				
The dynamical time is roughly equal to $\Delta H/c_{\rm s}$, 
where $\Delta H$ is the thickness of the multi-branch region, 
and $c_{\rm s}$ is the sound velocity in the cold phase. 
The sound velocity includes the radiation pressure, but 
in a proportion which depends on the relation between the 
radiation field and the gas. Since the multi-branch region  
is not very optically thick, one can neglect the radiation 
pressure, and simply assume $c_{\rm s}=\sqrt{2kT/m_{\rm H}}$, 
where $m_{\rm H}$ is the proton mass. The thickness of the 
multi-branch region is a fraction $f_{\rm multi}$ of the 
total thickness of the slab, which itself depends on the 
density (since the column density is a given parameter). 
Expressing the total column density in 10$^{23}$ cm$^{-2}$, 
one finds:

\begin{equation}
t_{\rm dyn} \sim 2\ 10^4\,{ f_{\rm multi} \, N_{23}\over 
\sqrt{T_5} \, n_{12}} \ \ \ \ {\rm (in~units~of~s)}. 
\label{tdyn} 
\end{equation}

It is interesting to note that all timescales depend on the same way 
on the density, which is often an unknown parameter when dealing 
with observational data (e.g., Warm Absorber gas). If the timescales 
vary for different locations in the slab (because the 
density value is not the same at different depths), their 
ratio does not change much. In particular, the ratio  
$t_{\rm dyn}/t_{\rm th}$  is very large (of the order of 
10$^3$ if the thermal instability affects 10\%\ of the slab), 
therefore the multi-branch region should contain local thermal 
perturbations out of pressure equilibrium. 

Chevallier et al. (2006b) made a detailed study on the viability 
of pressure equilibrium media  in the presence of variations 
of the incident radiation flux. The authors have shown that, though 
the global pressure equilibrium is generally preserved (with   
the medium responding to the variations by rapidly adjusting  
its ionization and thermal equilibrium), supersonic velocities 
develop in the gas, owing to local over- or under-pressure 
states. The same phenomenon should occur in the multi-branch region  
of our media, even in the absence of incident flux variations. 
To study this problem, a complete perturbational study should 
be performed around the equilibrium state of the gas slab;  
such a study is out of the scope of the present paper. For the 
time being, we can just conclude that large spectral fluctuations 
corresponding to the onset of a ``cold'' or a ``hot'' solution 
could be observed in timescales of the order of the dynamical 
time, independently of any variation of the radiation field 
external to the medium. Moreover, a strong turbulence implying 
supersonic velocities leading to line broadening should 
permanently exist in the multi-branch region of thick, stratified, 
pressure equilibrium media.	
	
\subsection{Practical considerations}
An important point to take into account when choosing 
which computational method to apply, is that the full 
computation of the ``hot'' and ``cold'' models described here 
is extremely time-consuming; this is because the process is strongly 
unstable and requires thus more iterations than the isodensity 
scheme models. Given the results presented in this paper, it 
seems thus reasonable to use the simpler, isodensity scheme to 
compute constant pressure models or hydrostatic equilibrium models.  
This procedure is less ad-hoc than to choose arbitrarily between 
one of the possible solutions  resulting, in the 
end, in gas structures and emitted/absorbed spectra very 
close to what is expected from an ionized medium consisting in 
similar proportions of gas in the hot and cold phases.

\section{Conclusions}\label{conclusions}

\begin{enumerate}

\item We have addressed the thermal instability issue in 
the case of optically thick, stratified media, in total 
pressure equilibrium. \\

\item In order to do that, we have developped and implemented 
a new algorithm in the \titan\ code; this algorithm, based on 
an isobaric computational method, allows to select the hot/cold 
stable solutions and thereof to compute a fully consistent photoionized 
model for each solution. \\

\item We have chosen to work with a few models encompassing 
the range of conditions valid for the Warm Absorber in AGN; 
our results can be applied to media in any pressure equilibrium 
conditions, e.g., constant gas pressure, constant total 
pressure, or hydrostatic pressure equilibrium. \\

\item We have shown that a thick, stratified medium ionized 
by X-rays behaves differently from a thin ionized medium.   
This happens for mainly two reasons:  first, the spectral 
distribution of the mean intensity at the illuminated surface 
of the slab is different from the incident spectrum, as 
it equally contains a ``returning" radiation component  
emitted by the slab itself; second, the spectral distribution 
changes as the radiation progresses inside the medium and,  
as a consequence, the shape of the S-curve also changes. 
These effects depend on the thickess of the medium and 
on its ionization. \\

\item This has observational implications in the emitted/absorbed 
spectra, ionization states, and variability. It is impossible 
to know what solution the plasma will adopt when attaining 
the multi-solutions regime:  it can  oscillate between the hot and
cold  solutions, it can fragment into hot and cold clumps 
which will coexist together, or it can take the form of 
a hot, dilute medium confining cold, denser clumps. Nevertheless, 
one expects the  emitted/absorbed spectrum  to be intermediate 
between those resulting from pure cold and hot models. \\

\item We have compared the results obtained with models based 
on the pure hot/cold solutions, and models computed with an 
approximate, intermediate solution; we have demonstrated that 
the pure hot/cold models represent two extreme results corresponding 
to a given gas composition and photoionizing flux, being compatible 
with the two stable solutions. 
The three (hot, cold, and intermediate) models differ not 
only in the layers where multiple solutions are possible, but 
all along the gas slab, this because the entire radiation field 
suffers modifications while crossing a thick medium. \\

\item The spectra emitted or absorbed by a given ionized medium, 
consisting of a mixture of gas in the hot and cold phases,  
should thus be intermediate between those resulting from the pure 
cold and hot models; therefore, the intermediate model provides 
a good description of such a mixed-phase medium. The 
differences between the emitted and absorbed spectra obtained 
with the stable solutions provide an indication for the 
maximum  ``error bars'' associated to the spectra computed 
with the ``intermediate'' solution. \\

\item The relative proportion of the hot and cold phases 
could vary with time. One expects two types of variability: 
a relativelly weak variability related only to a temperature 
variation, and a stronger variability related to a change of 
ionization state. Large spectral fluctuations corresponding 
to the onset of a cold/hot solution could be observed in 
timescales of the order of the dynamical time. Moreover, 
a strong turbulence implying supersonic velocities leading 
to line broadening should permanently exist in the 
multi-branch region of thick, stratified, pressure equilibrium 
media.	

\end{enumerate}

\vspace{1cm}
\begin{acknowledgements}
We acknowledge grant BPD/11641/2002 of the FCT, Portugal;  
grant 1P03~D00829  of the PSCSR, Poland, and  support from  
LEA and astro-PF, Poland-France. The authors acknowledge 
B. Czerny and A. \agata\ for fruitfull discussions on the 
subjects of thermal instability and electron conductivity.  
\end{acknowledgements}

\end{document}